\documentclass[aps,prb,twocolumn,longbibliography,superscriptaddress,floatfix]{revtex4-2}

\usepackage{graphicx}% Include figure files
\usepackage{dcolumn}% Align table columns on decimal point
\usepackage{bm}% bold math
\usepackage{amsmath}
\usepackage{amssymb}
\usepackage{amsthm}
\usepackage{mathtools}
\usepackage{mathrsfs}
\usepackage{float}
\usepackage{ulem}
\usepackage{epstopdf}
\usepackage{subfigure}
\usepackage{color}
\usepackage{newlfont}
\usepackage{epstopdf}
\usepackage{appendix}
\usepackage[breaklinks=true]{hyperref}
\usepackage{breakcites}
\usepackage{textcomp}
\usepackage{appendix}
\hypersetup{
     colorlinks   = true,
     linkcolor    = red,
     citecolor    = cyan,
     urlcolor     = magenta
     }

\begin{document}

% \preprint{APS/123-QED}

\author{Bertin Many Manda}
% \email{bertin.many\_manda@univ-lemans.fr}
\affiliation{Laboratoire d’Acoustique de l’Universit\'e du Mans (LAUM), UMR 6613, Institut d'Acoustique - Graduate School (IA-GS), CNRS, Le Mans Universit\'e, France}

\author{Ricardo Carretero-Gonz{\'a}lez}
% \email{rcarretero@sdsu.edu}
\affiliation{Nonlinear Dynamical Systems
Group,
% \footnote{\texttt{URL}: http://nlds.sdsu.edu}
Computational Sciences Research Center, and
Department of Mathematics and Statistics,
San Diego State University, San Diego, California 92182-7720, USA}

\author{Panayotis G. Kevrekidis}
% \email{kevrekid@math.umass.edu}
\affiliation{Department of Mathematics and Statistics, University of Massachusetts, Amherst, MA 01003-4515, USA}

\author{Vassos Achilleos}
% \email{achilleos.vassos@univ-lemans.fr}
\affiliation{Laboratoire d’Acoustique de l’Universit\'e du Mans (LAUM), UMR 6613, Institut d'Acoustique - Graduate School (IA-GS), CNRS, Le Mans Universit\'e, France}

% \preprint{APS/123-QED}

\title{Skin modes in a nonlinear Hatano-Nelson model}
%On nonlinear skin modes in the Hatano-Nelson lattice model: \\
%Existence, Stability and Dynamics}% Force line breaks with \\
% \thanks{A footnote to the article title}%

\date{\today}% It is always \today, today,
             %  but any date may be explicitly specified

\begin{abstract}
Non-Hermitian lattices with non-reciprocal couplings under open boundary conditions are known to possess linear modes exponentially localized on one edge of the chain. 
This phenomenon, dubbed non-Hermitian skin effect, induces all input waves in the linearized limit of the system to unidirectionally propagate toward the system's preferred boundary.
% has been extensively studied in various linear models supporting waves.
% reveling interesting localization properties. 
Here we investigate the fate of the non-Hermitian skin effect in the 
presence of Kerr-type nonlinearity within the well-established Hatano-Nelson lattice model.
Our method is to probe the presence of nonlinear stationary modes which are localized at the favored edge, when the Hatano-Nelson model deviates from the linear regime.  
Based on perturbation theory, we show that families of nonlinear skin modes emerge from the linear ones at any non-reciprocal strength.
Our findings reveal that, in the case of focusing nonlinearity, these families of nonlinear skin modes tend to exhibit enhanced localization, bridging the gap between weakly nonlinear modes and the highly nonlinear states 
(discrete solitons) when approaching the anti-continuum limit with vanishing couplings. Conversely, for defocusing nonlinearity, these nonlinear skin modes tend to become more extended than their linear counterpart.
To assess the stability of these solutions, we conduct a linear stability analysis across the entire spectrum of obtained nonlinear modes and also explore representative
examples of their evolution dynamics.
\end{abstract}

%\keywords{Suggested keywords}%Use showkeys class option if keyword
                              %display desired
\maketitle

%\tableofcontents
\section{\label{sec:intro}Introduction}

Advances in the studies of non-conservative systems using non-Hermitian operators have led  to the discovery of various   interesting phenomena. At the origin of these research activities lies the seminal work on $\mathcal{PT}$-symmetry ~\cite{BB1998,BBJ2002} describing systems 
featuring the simultaneous balance of energy dissipation and gain.
$\mathcal{PT}$-symmetry provides the means to construct non-Hermitian operators,
which under certain conditions may support modes with real 
eigenvalues~\cite{B2007,M2011}.
These results were confirmed and validated in many different physical domains including optics, acoustics, mechanical, and electrical systems; for a recent review, see e.g., the book~\cite{CY2018}. 
More recently, $\mathcal{PT}$-symmetry, was also employed in the context of topology and now plays a significant role in the studies of non-Hermitian topology~\cite{ZZCLC2023,KSUS2019}. It is now understood that non-Hermitian topology may substantially differ from its Hermitian counterpart.

Within the context of non-Hermiticity and topology, a new class of models featuring
asymmetric (often non-reciprocal) couplings between their constituents have been introduced.
Unlike Hermitian or $\mathcal{PT}$-symmetry systems, it is found that the spectral characteristics of systems with asymmetric (non-reciprocal) couplings greatly vary depending on their boundaries~\cite{YW2018,CH2021,CZLC2022,WC2023,ZZLC2022,LTLL2023,OS2023}.
In general, these type of non-Hermitian systems
under periodic boundaries possess complex spectra with Bloch-type linear modes.
Nevertheless, the same system with open boundary conditions (OBC) may display a real spectrum. More importantly, the eigenmodes under OBC are found to be exponentially localized on one side of the system.     
This phenomenon is dubbed the non-Hermitian skin effect (NHSE) and constitutes one of the latest developments in this area;
see e.g., Refs~\cite{LTLL2023,WC2023} for a review. In practice, the skin effect is a manifestation of the asymmetry of the couplings favoring the accumulation of amplitude in one side.
Note that most of the properties of the NHSE and of non-Hermitian topology can be understood  using the seminal linear Hatano-Nelson (HN) model~\cite{HN1996,HN1998}, a non-Hermitian one-dimensional lattice with asymmetric nearest neighbor couplings.
The intense interest in this structures has led to several experimental manifestations of the NHSE in optics~\cite{WKHHSGTS2020}, acoustics~\cite{ZYGGCYCXLJYSCZ2021,MAPPA2023}, mechanics~\cite{BLC2019,GBVC2020}, electric circuits~\cite{LSMZYWJJZ2021}, and atomic lattices~\cite{LXDLLGYY2022}.

As is often the case, the  phenomena of non-Hermitian and topological systems have been pursued also in the realm of nonlinear waves. On one hand, the study of the interplay between nonlinearity and topology is rapidly expanding, giving rise the new phenomena such as topological breathers and solitons~\cite{LC2016,FF2019,CXYKT2021,JD2022,K2023,NYKA2023}. In addition, the interaction between non-Hermiticity and nonlinearity, has been extensively focused mainly on $\mathcal{PT}$-symmetric systems (see, 
e.g., the reviews~\cite{KYZ2016,SSHDLK2016,CY2018}).
Even more recently the experiments exploiting the inteperlay between topology, $\mathcal{PT}$-symmetry and nonlinearity have been performed~\cite{sciencemakris21}.

To our knowledge, models featuring asymmetric (non-reciprocal) interactions and the NHSE have been merely studied under the effect of nonlinearity. In fact only few works~\cite{Y2021,E2022,ZWLXWC2022,JCZL2023} recently appeared with Kerr-type nonlinearities.  
In particular, Refs.~\cite{Y2021,JCZL2023} focus on a few site lattice, and at the extreme limit where each lattice site is only connected to its right (or left) nearest neighbor.
In addition, Ref.~\cite{E2022} studied the dynamics of single-site wave-packets at the center of the HN chain, demonstrating that nonlinear NHSEs exist, but may be hindered by self-trapping processes.

Here, we intend to shed more light on the fate of the NHSE of the HN model in the nonlinear regime. To do so we study the HN model with Kerr-type nonlinearity ensuing from  a non-reciprocal variant of the discrete nonlinear Schr\"{o}dinger (DNLS) equation~\cite{K2009}. We thus focus on finding nonlinear skin modes (NLSMs) stemming from the corresponding linear ones. By applying perturbation theory we show that small amplitude NLSMs emerge for all linear modes both with focusing (positive) or defocusing (negative) signs of the nonlinearity. 
We find families of NLSMs which connect the linear and the strongly nonlinear anticontinuum (AC) limit where the coupling between each site vanishes. 
We also carry a linear stability analysis and identify regions where NLSMs can be either stable or unstable. Our analysis shows that small perturbations that do not have a skin-like profile will eventually destabilise the NLSMs.

The paper is structured as follows. In Sec.~\ref{sec:model_&_skinmodes}, we discuss the existence of such NLSMs. In Sec.~\ref{sec:stability_and_dynamics}, we tackle their stability and
dynamics. Finally in Sec.~\ref{sec:concl}, we present our conclusions and a number
of directions for future studies. Finally, in the Appendices we provide a 
number of details regarding our perturbation theory analysis in the
different limits.

%%%%%%%%%%%%%%%%%%%%%%%%%%%%%%%%%%%%%%%%%%%%%%%%%%%%%%%%%%%
\section{\label{sec:model_&_skinmodes}Nonlinear skin modes}
%%%%%%%%%%%%%%%%%%%%%%%%%%%%%%%%%%%%%%%%%%%%%%%%%%%%%%%%%%%

\begin{figure}[t]
    \centering
    \includegraphics[width=\columnwidth]{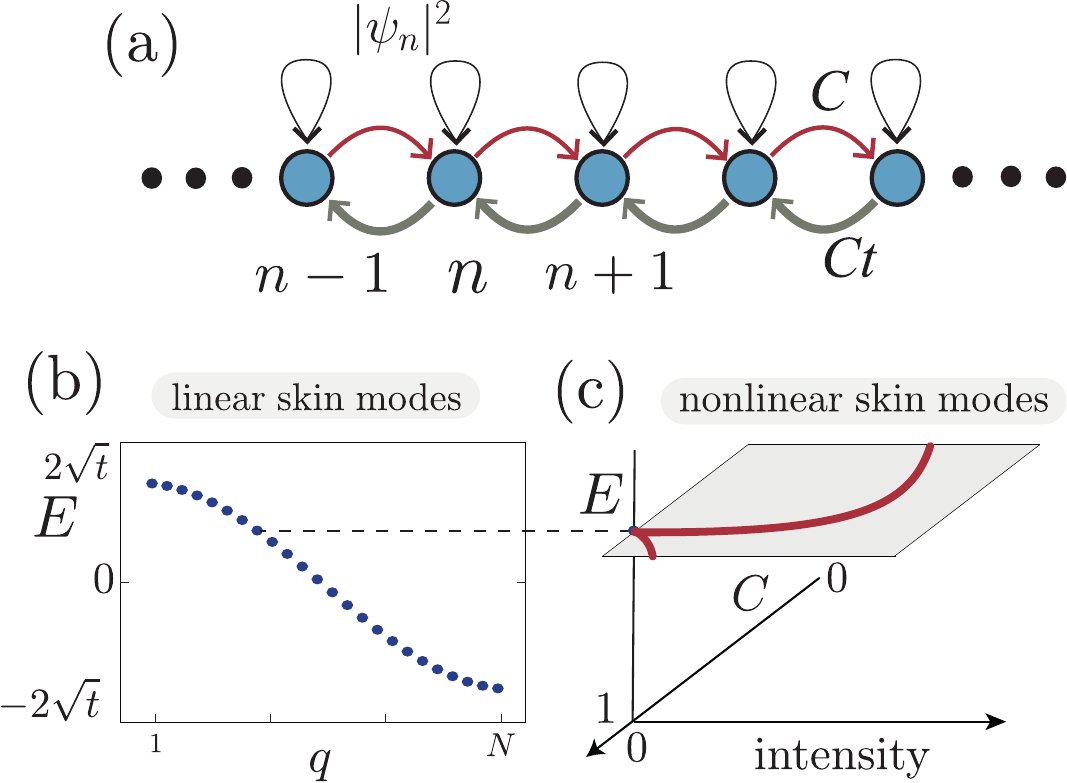}
    \caption{(a) The proposed nonlinear HN model where closed loops represent the  on-site Kerr terms.
        (b) The linear spectrum of a HN model with with open boundary conditions. (c) An illustration of how a family of nonlinear skin modes emerges from the corresponding linear solution.
        % The $\psi_n$ is complex amplitude at site $n\in \mathbb{Z}$.
    }
    \label{fig:HNLattice}
\end{figure}

In this work we are interested in finding nonlinear skin modes  of the following nonlinear version of the HN model
\begin{equation}
    i\frac{d\psi _n}{d\tau} = C \left(\psi_{n+1} + t\psi_{n-1}\right)+\sigma\lvert \psi_n\rvert ^2 \psi_n,
    \label{eq:dynHN_model}
\end{equation}
where $n=1, 2, \ldots, N$ indexes the lattice sites and $\psi_n$ is a 
complex-valued amplitude at site $n$,  as shown in Fig.~\ref{fig:HNLattice}(a).
Here, $C$ is the (real) hopping (coupling) strength and $t$ is the non-reciprocal parameter: if $t=1$ the system is reciprocal while for any value $t\ne 1$ there is a preferred direction. 
We further assume a general Kerr-type (self-interaction) nonlinearity with strength $\sigma = \pm 1$ relevant to many  physical systems especially  in nonlinear optics~\cite{CJ1988,ESMBA1998,LEDERER20081} and cold atomic gases~\cite{L2001,RevModPhys.78.179}, or even bio-molecules~\cite{S1992,niemi}.  Indeed, this model in the $t=1$ limit constitutes the prototypical nonlinear
dispersive dynamical lattice of the DNLS
type~\cite{K2009}. Here, we explore a hybrid between this well-established 
setting and the linear HN model, to examine the interplay of asymmetric
dispersion and cubic nonlinearity.

%%%%%%%%%%%%%%%%%%%%%%%%%%%%%%%%%%%%%%%%%%%%%%%%%%%%%%%%%%%
%%%%%%%%%%%%%%%%%%%%%%%%%%%%%%%%%%%%%%%%%%%%%%%%%%%%%%%%%%%
 Our goal is to find  stationary solutions, with energy $E$, assuming $\psi_n(\tau) = u_n e^{iE\tau}$ which leads to the following set of nonlinear equations~\cite{JCZL2023,Y2021}
\begin{equation}
    Eu_n = C \left(u_{n+1} + tu_{n-1}\right)+\sigma\lvert u_n\rvert ^2 u_n.
    \label{eq:HNNL_model}
\end{equation}
The linear HN model  (obtained for $\sigma=0$ and $C=1$) yields
\begin{equation}
    Eu_n = u_{n+1} + tu_{n-1}.
    \label{eq:static_linear_hn_model}
\end{equation}
For a finite HN lattice with OBC  $u_0=u_{N+1}=0$, Eq.~\eqref{eq:static_linear_hn_model} can be recast in the form of an eigenvalue problem $H \vec{u}_q^{(0)}=E_q\vec{u}_q^{(0)}$ (the expression of $H$ is provided in Appendix~\ref{sec:app:matrices}), where the real eigenenergies are equal to 
\begin{equation}
    E_q = 2\sqrt{t}\cos \left(\frac{q\pi}{N+1} \right),
    \label{eq:eigenenergy}
\end{equation} 
with  $q=1,\ldots,N$. 
An example of such a spectrum is shown in Fig.~\ref{fig:HNLattice}(b) for $N=24$.
The corresponding eigenvectors $\vec{u}_q^{(0)}$ have elements which satisfy the following equation
\begin{equation}
    u^{(0)}_{q,n}=\sqrt{\frac{2}{N+1}}t^{n/2}\sin \left(\frac{nq\pi}{N+1} \right).
    \label{eq:skineigs}
\end{equation}
From Eq.~\eqref{eq:skineigs}, it becomes clear that whenever $t<1$ ($t>1$) the modes of the HN are localized to the left- (right-) hand side of the lattice owing to the $t$-dependent prefactor.
This is exactly the manifestation of the NHSE which we systematically 
generalise here in the nonlinear domain. In fact, since the eigenvalue problem is non-Hermitian, there exist also left eigenvectors, $\vec{v}_q$, satisfying 
$\vec{v}_q^{T(0)}H =\vec{v}_q^{T(0)}E_q$. 
These left eigenvectors are localized at the opposite side of the lattice. 
It is worth noting that we normalize these eigenvectors using the bi-orthonormalization, i.e., $\vec{v}_{q'}^{T(0)}\vec{u}_q^{(0)}=\delta _{q'q}$~\cite{EKS2023}.

Our main goal is to show that  families of NLSMs satisfying Eq.~\eqref{eq:HNNL_model} emerge from their linear counterparts for our finite lattices. In particular, we choose to keep the energy $E$ (i.e., the nonlinear eigenvalue parameter,
referred to as propagation frequency in optics and chemical potential
in atomic physics) fixed to that of the linear modes, and then numerically obtain solutions by varying the coupling strength, $C$. 
A sketch of this procedure and its outcomes is shown in 
Fig.~\ref{fig:HNLattice}(c).
Let us first use the regular perturbation theory (RPT) to show that 
for any value of $t\ne 0$, small amplitude nonlinear solutions can emerge from each linear mode. 
To this end, we assume solutions of Eq.~\eqref{eq:HNNL_model} in the following asymptotic form
\begin{equation}
    u_n =  \epsilon u_n^{(0)} + \epsilon^2 u_n ^{(1)} + \epsilon^3 u_n^{(2)} + \mathcal{O}\left(\epsilon^4\right),
\end{equation}
where $u_n^{(0)}$ are the linear modes [Eq.~\eqref{eq:static_linear_hn_model}]. 
Since the continuation is performed at constant energy, $E_q$ for each of the modes, 
$C$ then becomes the continuation parameter that varies in a way such that for sufficiently small amplitude solutions we consider
\begin{equation}
    C = 1 + C_1 \epsilon^2 + \mathcal{O}\left(\epsilon^4\right).
\end{equation}
The details of the perturbation theory are found in the Appendix~\ref{sec:app:rpt_linear}; here we only present the main results. 
We find that finite amplitude solutions of the nonlinear eigenvalue problem [Eq.~\eqref{eq:HNNL_model}] exist starting from terms of $\mathcal{O}\left(\epsilon^3\right)$ with a correction to $C$ given by
\begin{equation}
    C_1 =-\frac{\sigma}{E_q}\left(\frac{2}{N+1}\right)^2\sum_{n=1}^Nt^{n}\sin^4 \left(\frac{nq\pi}{N+1} \right).
    % C_1 =-\frac{\sigma}{E_q}\left(\frac{2}{N+1}\right)^2\sum_{n=1}^Nt^{n}\sin^4 \left(\frac{nq\pi}{N+1} \right).
    \label{eq:c1}
\end{equation}
We note that this expression can be resummed, provided that we decompose
the $\sin^4$ into Fourier modes and consider the resulting geometric series,
yet given the intricacy of the resulting expression, we do not explore that
step herein.
In practice, the above result says that  for small amplitudes a family of nonlinear modes emerges from each respective linear mode followed by a change in the coupling parameter $C$. 
Furthermore the negative sign in front of Eq.~\eqref{eq:c1} signals the fact that for $\sigma=+1$ (focusing case) the coefficient $C$ decreases as the amplitude grows, while the opposite happens for the defocusing case. In addition we see that the correction $C_1$ strongly depends on the value of the non-reciprocal parameter $t$.

\begin{figure}
    \centering
    \includegraphics[width=\columnwidth]{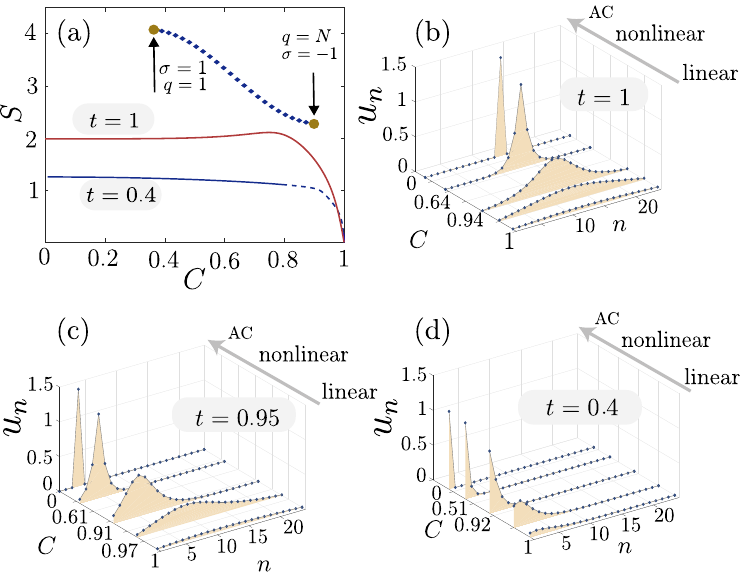}
    \caption{
        (a) The total intensity, $S$, as a function of the coupling strength, $C$, for the family of nonlinear stationary solutions emerging from the first linear mode with $t=1$ (red curve) and $t=0.4$ (blue curve).
        The dashed regions of the curves indicate linearly unstable solutions.
        Further, the inset shows the linear spectrum.
        (b) Representative nonlinear modes of the family with $t=1$ for different $C$ values. (c) and (d) Same as (b) but for the families of NLSMs with $t=0.95$ and $t=0.4$, respectively. 
        % In panel (d) the AC solution is localized only on the first site.
    }
    \label{fig:continuation1}
\end{figure}
%%%%%%%%%%%%%%%%%%%%%%%%%%%%%%%%%%%%%%%%%%%%%%%%%%%%%%%

%%%%%%%%%%%%%%%%%%%%%%%%%%%%%%
\subsection{Focusing regime}
%%%%%%%%%%%%%%%%%%%%%%%%%%%%%%

To verify the above result, we solve Eq.~\eqref{eq:HNNL_model} numerically by fixing $E$ to a linear eigenvalue $E_q$ [Eq.~\eqref{eq:eigenenergy}] 
 employing a pseudo-arclength nonlinear solver~\cite{DKK1991,DGKMS2008,matcont2023}. 
The initial guess seeded to the solver is the corresponding  linear mode rescaled to very small amplitudes. 
That way the value of $C$ is left as an unknown parameter to be found by the pseudo-arclength solver. 
In Fig.~\ref{fig:continuation1} we show results for the case of the first linear mode ($q=1$) for a lattice of $N=24$ oscillators. 
Figure~\ref{fig:continuation1}(a) shows the numerically calculated total intensity of the lattice 
\begin{equation}
    S=\sum_{n}\lvert u_n\rvert^2,
    \label{eq:total_intenstiy}
\end{equation}
as a function of the coefficient $C$ for two values of $t$ mapping to the usual 
DNLS equation with $t=1$ and the nonlinear HN model with $t=0.4$.
As predicted, a family of nonlinear modes emerges from the linear limit with decreasing values of $C$. 
In addition, beyond the validity of the perturbation series in the neighborhood of the linear regime, we numerically find that each family terminates at the AC limit, where the oscillators are uncoupled, i.e., $C=0$.
% with a single excited site. 
In fact, the existence of this branch bifurcating from the AC limit can also be demonstrated through regular perturbation theory from the latter limit. 
The details of these calculations are given in Appendix~\ref{sec:app:RPTAClimit}.

For illustrative purposes, in Fig.~\ref{fig:continuation1}(b) we show the profile of some solutions for the family with $t=1$ (corresponding to the DNLS model). 
Clearly, small amplitude nonlinear solutions are similar to the well-known sinusoidal form of the linear mode in the neighborhood of unit coupling strength.
As $C$ decreases away from unity and nonlinearity becomes stronger, these solutions become more localized yet remain spatially symmetric (i.e. the location of their centers of 
mass does not change) with in-phase amplitudes at adjacent oscillators. 
At the AC limit ($C=0$), the corresponding family of modes is connected to a single site solution at the center of the lattice satisfying $E_1=\lvert u_{N/2}\rvert^2$~\cite{BIFURCT1}. 
Note that this is a finite size effect in contrast to the well known results 
for $N\rightarrow \infty$ where  single-site solutions in the AC limit are 
connected to the continuum soliton, as the coupling is increased~\cite{K2009}.

Moving away from the DNLS limit, in Fig. \ref{fig:continuation1}(c), we display instances of NLSMs of the HN model with 
 $t=0.95$. 
 As expected small amplitude solutions remain close to the shape of the linear skin mode. 
However as the amplitudes grows ($C$ decreases) the center of mass of the obtained modes moves toward the favored direction [for $t<1$, i.e., 
to the left side of the chain of Fig.~\ref{fig:HNLattice}(a)]. 
It follows that focusing nonlinearity not only tends to localize the modes but also strengthens the skin effect. 
In fact, even for a small deviation from the standard DNLS model in this example ($t=0.95$), the resulting family of NLSMs connects to the single-site solution located far away from the lattice's center as seen in Fig.~\ref{fig:continuation1}(c).
More importantly, we numerically confirmed that this shifting toward the left 
becomes stronger as $t$ is further decreased, till the families of NLSMs starting from the first linear mode, end up at a single-site solution in the AC limit, located at the left edge of the chain.
This happens for all values of $t$ below the threshold, $t_c \approx 0.57$.
Interestingly the $t_c$ happens to be independent of lattice size, at least up to the largest chain ($N=50$) considered in our numerical simulations. 
Examples of such a family of NLSMs is shown in Fig.~\ref{fig:continuation1}(d) for $t=0.4$.
\begin{figure*}
    \centering
    \includegraphics[width=0.9\textwidth]{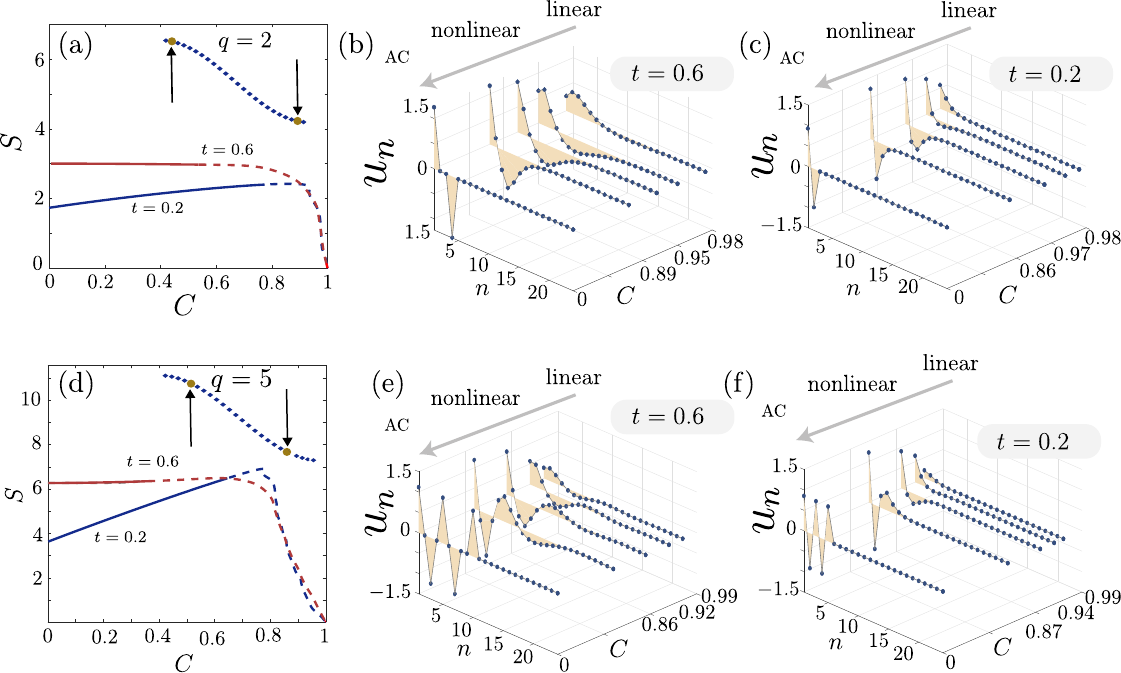}
    \caption{
        (a) Intensity $S$ as a function of $C$ for the family of NLSMs originating from the second linear mode at two different values of $t$ with $t=0.2$ and $t=0.6$. Note that solid vs.~dashed reflects the stable vs.~unstable branches; also, note that additional branches to the ones originating from the linear limit are depicted in this panel (see details in the text).
        (b) Representative profile of the NLSM associated with the family originating from the second linear mode with $t=0.6$, red curve in (a).
        (c) Same as in (b), but for $t=0.2$, blue curve in (a).
        (d) Same as (a), but for the families of NLSMs emerging from the fifth linear mode.
        (e) Same as (b) but for (d).
        (f) Same as (f) but for (d).
        }
    \label{fig:continuation2}
\end{figure*}
We note in passing the staggering transformation $u_n \rightarrow (-1)^nu_n^\prime$ within 
Eq.~\eqref{eq:HNNL_model}, leads to  the following  
\begin{equation}
    -Eu_n^\prime = C \left(u_{n+1}^\prime + tu_{n-1}^\prime\right)-\sigma\lvert u_n^\prime\rvert ^2 u_n^\prime.
    \label{eq:srtagg}
\end{equation}
Consequently, if ($E$, $u_n$, $\sigma$) is a solution of the nonlinear eigenvalue problem, then ($-E$, $u_n^\prime$, $-\sigma$) is also a solution. 
As such the families of NLSMs generated through the linear mode of wave number $q^\prime=N+1-q$ with $\sigma = -1$ overlaps with the ones arising from the linear mode with index $q$, fixing $\sigma=1$; see the inset of Fig.~\ref{fig:continuation1}(a).

Let us now look for families of NLSMs stemming from linear modes of higher wave number. 
The $S$ vs.~$C$ plot resulting from the numerical continuations is shown in the panels of Fig.~\ref{fig:continuation2} using as initial guess the linear modes of wave numbers $q=2$ [Fig.~\ref{fig:continuation2}(a)] and $q=5$ [Fig.~\ref{fig:continuation2}(e)], considering two different values of the strength of non-reciprocity with $t=0.2$ (blue curves) and $t=0.6$ (red curves). 
Similarly to  Fig.~\ref{fig:continuation1} these families join the linear 
regime at $C=1$ with the AC limit of $C=0$. 
Focusing now on the shapes of the resulting NLSMs, we find that they inherit the number of nodes pertaining to the order of the mode in the linear counterpart.
Namely, the 2nd mode results in a 2-site state, the 5th mode to a 5-site one, etc.
Given their excited nature, the resulting NLSMs are also ``twisted'' 
(i.e., corresponding to alternating, out-of-phase, field values)~\cite{PKF2005,K2009}.
%While the location of these nodes is the same as for the linear mode for small amplitude NLSMs, 
These NLSMs also tend to shift toward the preferred direction of the system at high intensity, see e.g., Figs.~\ref{fig:continuation2}(b) and (c).
As such, the shape of the high amplitude NLSMs, greatly differs from the linear one.
In particular, for the examples in Figs.~\ref{fig:continuation2}(b) and (c) and Fig.~\ref{fig:continuation2}(e) and (f) the shape of these high amplitude NLSM becomes closer to the  
% However, as the amplitude increases, their shape gradually changes and they end up to a 
$2$-site and $5$-site stationary solutions of the AC limit, respectively for families emerging from the second [Figs.~\ref{fig:continuation2}(b) and (c)] and fifth [Figs.~\ref{fig:continuation2}(e) and (f)] linear modes.
It is important to emphasize that from our simulations, we conjecture that the $q$-th linear mode of unit coupling is connected to a $q$-site solution in the AC limit, with out-of-phase excited oscillators. The latter, as was illustrated
for $t=1$ in Ref.~\cite{PKF2005} and  as follows from the 
calculations in Appendix B for $t \neq 1$, 
are spectrally stable states near the AC limit.

Interestingly, the location of the excited oscillators at the 
AC limit depends on the strength of the non-reciprocity, $t$.
Namely, at the $t=1$ (DNLS) limit, we expect the positions of 
the excited oscillators to be equally spaced inside the bulk of the chain due to the symmetry of the model.
However, when $t\neq 1$, we have seen in Fig.~\ref{fig:continuation1}, that 
the NHSE tends to shift the positions of these excited oscillators {\it independently} toward the most favorable direction of wave propagation within the lattice.
This can be clearly seen when comparing the families emerging from the second linear modes at $t=0.6$ and $t=0.2$ respectively in Fig.~\ref{fig:continuation2}(b) and  Fig.~\ref{fig:continuation2}(c).
The former features excited oscillators with indices $n=1$ and $n=4$ [Fig.~\ref{fig:continuation2}(b)], while the latter ones with $n=1$ and $n=2$ [Fig.~\ref{fig:continuation2}(c)] at $C=0$.  
A similar observation can be drawn from Figs.~\ref{fig:continuation2}(e) and~\ref{fig:continuation2}(f),
where the larger $t$ can afford the separation of the excited sites by
empty ``holes'', while for the lower $t$, these holes are suppressed in favor
of the adjacent excitation of all the nonzero nodes at the AC limit.
Furthermore, we computed the critical strength of non-reciprocity, $t_c$ below which families of NLSMs emerging from the linear modes connect to consecutive excited oscillators in the AC limit.
We obtain that $t_c$ is independent on the lattice size, and has values $t_c \approx 0.25$ for $q>1$, in line with the results of Fig.~\ref{fig:continuation2}.

\begin{figure*}
    \centering
   \includegraphics[width=0.9\textwidth]{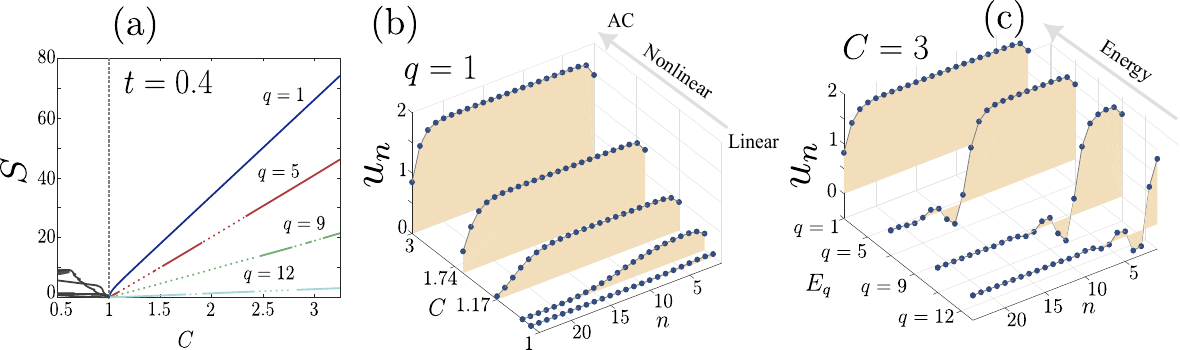}
    \caption{(a) Total intensity, $S$ against $C$ for the full families of stationary states emerging from the linear modes with $q=1$, $q=5$, $q=9$, and $q=12$ at constant $E$ for $t=0.4$.
    Dotted regions represent unstable NLSMs.
    (b) Five representative NLSMs of the family emerging from the first linear mode at different coupling strengths.
    (c) Same as (b), but at fixed value of $C$ with different energies. 
    }
    \label{fig:continuation_defocusing_e123}
\end{figure*}

%%%%%%%%%%%%%%%%%%%%%%%%%%%%%%%%%%%%%%%%%%%%%%%%%%%
\subsection{\label{subsec:defocusing}Defocusing regime}
%%%%%%%%%%%%%%%%%%%%%%%%%%%%%%%%%%%%%%%%%%%%%%%%%%%

It is also worth discussing the defocusing case ($\sigma=-1$).
According to the regular perturbation theory and especially Eq.~\eqref{eq:c1}, 
families of NLSMs  with increasing $C$ can emerge from the linear limit with defocusing 
nonlinearity, i.e., for $\sigma=-1$.
This is confirmed by our numerical simulations.
In Fig.~\ref{fig:continuation_defocusing_e123}(a) we depict the total intensity as a function of the coupling strength for selected numbers of families of NLSMs arising from the linear modes with index $q=1$, $q=5$, $q=9$ and $q=12$ for the case $t=0.4$. 
For these solutions the total intensity $S$ grows monotonically with respect to $C$ without showing any sign of saturation, in contrast to what is seen in the focusing case;
see gray curves of Fig.~\ref{fig:continuation_defocusing_e123}(a).

Figure~\ref{fig:continuation_defocusing_e123}(b) depicts representative NLSMs of the family emerging from the first linear mode with $t=0.4$.
A direct comparison of these results can be carried with respect to the focusing case in Fig.~\ref{fig:continuation1}(d).
It follows that, contrary to what is seen in the focusing regime, 
in the defocusing case, the linear skin mode tends to widen as the total intensity of the system grows. This is in line with the defocusing nature of the nonlinearity
which apparently dominates the NHSE of the corresponding linear problem for the
$q=1$ state.
This increase in width of the NLSMs, leads to almost extended nonlinear states at high  amplitude.
Nevertheless, we find that such extended states are not reached for all linear mode energies.
In fact, the families of NLSMs arising from linear skin modes of higher wave number display clear localization at the edge of the chain even for high amplitude states. In fact, by fixing the value of $C$ and comparing NLSMs of increasing $q$ we find that their width tends to decay as shown for example in Fig.~\ref{fig:continuation_defocusing_e123}(c). Additionally,
an interesting feature of an oscillatory tail can be seen to develop for
this panel's case of $C=3$. As this regime is far from the regime accessible
to perturbative analysis, we do not pursue this feature further.

% \FloatBarrier
\section{\label{sec:stability_and_dynamics}Stability and dynamics}

For a complete study of the NLSMs we need to study their stability under the effect of small perturbations. Such a linear stability analysis is typical for the DNLS model and is usually performed substituting the ansatz
 $\vec{\psi}(\tau) = \left(\vec{u} + \epsilon \vec{w}(\tau)\right) e^{iE\tau}$ into Eq.~\eqref{eq:dynHN_model} with $\vec{u}$ being the numerically obtained NLSM 
 and $\epsilon\vec{w}$ a small perturbation. 
 By expressing the elements of the perturbation vector as   $w_n (\tau) = a_ne^{i\lambda \tau} + b^*_n e^{-i\lambda^\star \tau}$, where the asterisk  denotes 
 the complex conjugate, we end up with the following eigenvalue problem~\cite{PKF2005}  
\begin{equation}
{JL} \begin{pmatrix} \vec{a} \\ \vec{b} \end{pmatrix}=
    \begin{pmatrix} \vec{a} \\ \vec{b} \end{pmatrix} ,    \label{eq:stab_problem_01}
\end{equation}
where we represent the stability matrix $Z\equiv JL$ as a product of two matrices: $J$ the symplectic matrix and $L$ a linearization  matrix  which depends on the stationary state $u_n$ and acts on the spatial part of the perturbation eigenvector
$w_n$ (further details are in Appendix~\ref{sec:app:matrices}).
A NLSM is said to be linearly stable if the eigenvalues $\lambda$ of $Z$ are real. The matrix ${J}$ is skew symmetric while ${L}^T\ne{L}$ is generally non-Hermitian for $t\ne 1$. However by using the similarity transformation  $\widetilde{L}=D^{-1}LD$ we can show that $\widetilde{L}^T=\widetilde{L}$. Here, using $D=\mbox{diag}(d_0,d_1, \ldots,d_{N-1},d_0,d_1, \ldots, d_{N-1})$ with elements $d_n=\sqrt{t^n},\quad n=0,\ldots N-1$ allows us to 
 rewrite the stability eigenvalue problem [Eq.~\eqref{eq:stab_problem_01}] as follows 
\begin{equation}
    {J\widetilde{L}} 
    \begin{pmatrix} {\widetilde{\vec{a}}} \\ \widetilde{\vec{b}} \end{pmatrix}=
    \lambda 
    \begin{pmatrix} {\widetilde{\vec{a}}} \\ \widetilde{\vec{b}} \end{pmatrix}
    \label{eq:stab_problem_01tr},
\end{equation}
where $\widetilde{\vec{a}}=D^{-1}\vec{a}$ and $\widetilde{\vec{b}}=D^{-1}\vec{b}$. 
It follows that the matrix product $J\widetilde{L}$ is real and symplectic.
Thus the eigenvalues of the stability problem [Eq.~\eqref{eq:stab_problem_01}] come in quartets $\lambda$, $\lambda^*$, $-\lambda$ and $-\lambda^*$. 

%%%%%%%%%%%%%%%%%%%%%%%%%%%%%%%%%%%%%%%%%%%%%%%%%%%%%%%%%%%%%%%%%%%%%%%%%%%%%%%%%%
\begin{figure}[ht!]
    \centering\includegraphics[width=\columnwidth]{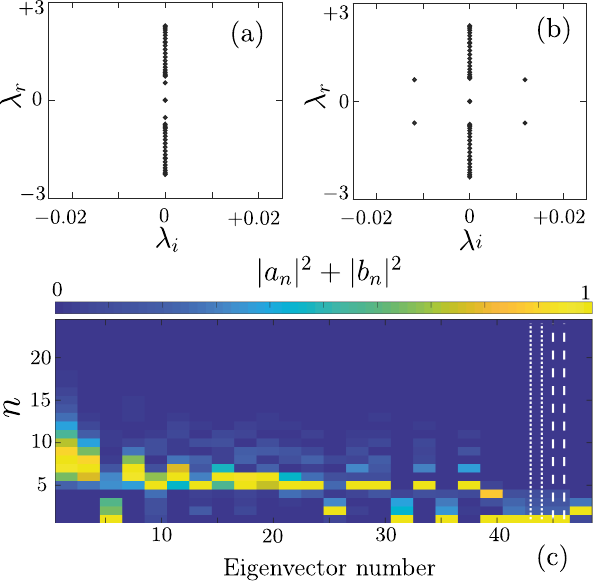}
    \caption{
    Eigenvalues $\lambda$ of the linear stability problem for the stationary solutions with (a) $C=0.501$ and 
    (b) $C=0.527$ of the family of NLSMs emerging from the second skin linear mode at constant energy, red curve in Fig.~\ref{fig:continuation2}(a).
    (c) Amplitude, $\lvert a_n\rvert^2 + \lvert b_n\rvert^2$,  of the eigenvectors of the eigenvalues in panel (b) sorted by decreasing participation number (see text for details).
    For clarity, the amplitudes are rescaled by the maximum at each eigenvector.
    The dotted and dashed line indicates the most unstable eigenvectors associated with eigenvalues with the largest positive and negative imaginary parts respectively [see panel(b)]. 
    }
    \label{fig:eigenvalues_eigenvectors_01a}
\end{figure}
%%%%%%%%%%%%%%%%%%%%%%%%%%%%%%%%%%%%%%%%%%%%%%%%%%%%%%%%%%%%%%%%%%%%%%%%%%%%%%%%%%

In Fig.~\ref{fig:continuation1}(a), Fig.~\ref{fig:continuation2}(a) and Fig.~\ref{fig:continuation_defocusing_e123}(a), the $S$ against $C$ dependence of the representative families of NLSMs also displays the results of their linear  stability analysis.
Namely, we mark stable regions by bold (colored) curves, otherwise unstable.
In the focusing case,  we find that the  NLSMs are stable in the vicinity of the linear regime as  depicted in Fig.~\ref{fig:RPT_Linear} of Appendix~\ref{sec:app:rpt_linear}.
The same applies in the neighborhood of the AC limit, whose stable nature is well shown in Fig.~\ref{fig:continuation1}(a), Fig.~\ref{fig:continuation2}(a) and Fig.~\ref{fig:RPT_AClimit}.
These two regions of stability typically sandwich unstable regions of NLSMs with moderate (and/or high) intensity [see e.g., Fig.~\ref{fig:continuation1}(a) and Fig.~\ref{fig:continuation2}(a)].

%%%%%%%%%%%%%%%%%%%%%%%%%%%%%%%%%%%%%%%%%%%%%%%%%%%%%%%%%%%%%%%%%%%%%%%%%%%%%%%%%%
\begin{figure*}
    \centering  \includegraphics[width=\textwidth]{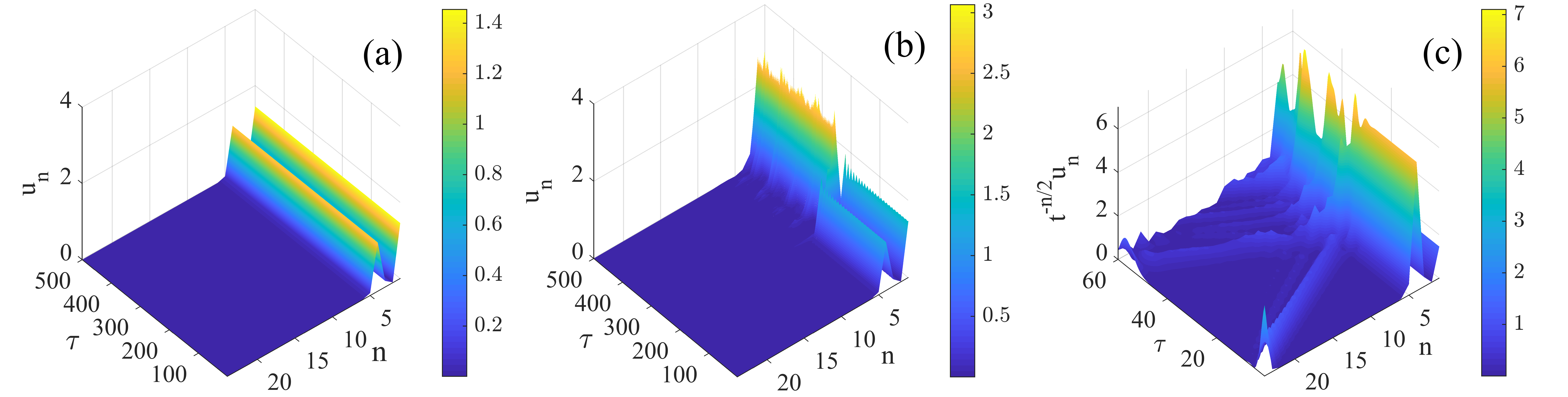}
    \caption{Nonlinear HN lattice dynamics.
        The initial conditions $\psi_n(\tau = 0) = u_n (\tau = 0) + \epsilon w_n (\tau = 0)$ with $\epsilon = 3\times 10^{-3}$ and $\lvert \vec{w} \rvert = 1$.
        (a) Stable NLSM, $u_n(\tau = 0)$, with $C=0.501$ and $t=0.6$ from the family 
        of the second ($q=2$) linear mode (see Fig.~\ref{fig:continuation2}) 
        with perturbation $w_n(\tau = 0) = a_n + b_n$ constructed with a random sum
        of eigenvectors of the stability matrix $Z=JL$ [Eq.~\eqref{eq:stab_problem_01}].
        (b) Unstable NLSM, $u_n(\tau = 0)$, with $C=0.527$ and $t=0.6$ belonging 
        to the same branch as in panel (a) with perturbation $w_n(\tau = 0)$ corresponding
        to the eigenvector of $Z$ along the most unstable direction.
        (c) Same as in (a), but for  
        $w_n(\tau = 0) = (0, 0, 0, \ldots, 0, e^{i\kappa_N})$, 
        with $\kappa_N = 2.727$ being randomly drawn from $[-\pi, \pi]$.
        Note that we have
        verified that the result of (c) is 
        representative of the dynamics of perturbations  with different realizations of the random phase.
        % randomly selected in the interval $[-\pi, \pi]$.
    }
    \label{fig:HNdynamics}
\end{figure*}
%%%%%%%%%%%%%%%%%%%%%%%%%%%%%%%%%%%%%%%%%%%%%%%%%%%%%%%%%%%%%%%%%%%%%%%%%%%%%%%%%%

Regarding the eigenspectrum of the stability matrix, representative cases are shown for a stable [Fig.~\ref{fig:eigenvalues_eigenvectors_01a}(a)] and complex (oscillatory) 
unstable [Fig.~\ref{fig:eigenvalues_eigenvectors_01a}(b)] NLSMs of the family emerging from the second linear mode with $t=0.6$, see the red curve in Fig.~\ref{fig:continuation2}(a).
The stable mode is taken at $C=0.501$ and the unstable one at $C=0.527$.
In addition, in Fig.~\ref{fig:eigenvalues_eigenvectors_01a}(c), we depict the eigenvectors associated with the eigenvalues of Fig.~\ref{fig:eigenvalues_eigenvectors_01a}(b) sorted by increasing value of the participation number $\left(\sum_k \lvert a_k\rvert^2 + \lvert b_k \rvert^2 \right)^2/\sum_n \left(\lvert a_n\rvert^2 + \lvert b_n \rvert^2 \right)^2$.
These eigenvectors come in pairs and display ``skinny''-like  profiles which 
%are reminiscent of the non-reciprocity within 
are a clear manifestation of the features of the HN model.
Furthermore, this can be
used to represent the generic form of a perturbation through which the linear stability of the NLSMs is defined.

To confirm the stability analysis, we numerically solve the equations of motion~\cite{HNW1993,DMMS2019,freelyDOP853} of the HN chain [Eq.(\ref{fig:HNLattice})] using an initial condition  of the form
$\vec{\psi}(\tau=0) = \vec{u} +\epsilon\vec{w}(\tau=0)$ where $u_n$ is the numerically obtained NLSM and $w_n(\tau=0) = a_n + b_n$ an initial deviation vector with unit norm.
Furthermore, we set $\epsilon = 3\times 10^{-3}$.
% is a pair of the obtained linear stability vectors and epsilon is a small parameter which we choose to fix to $\epsilon=1O^{-4}$.
The time evolution of the \textit{stable} NLSM of Fig.~\ref{fig:eigenvalues_eigenvectors_01a}(a) perturbed along the direction of one of the eigenvectors of its stability matrix is shown in Fig.~\ref{fig:HNdynamics}(a).
% perturbed by one of its linear stability eigenvectors; 
As expected the mode preserves its shape throughout the whole numerical integration.
Repeating the same procedure with the unstable NLSM in Fig.~\ref{fig:eigenvalues_eigenvectors_01a}(b) of the same family, eventually we observe that it deforms its shape as shown in Fig. \ref{fig:HNdynamics}(b). 
In fact, we see that after the instability sets in, the total intensity of the field grows substantially, while the state remains localized at the edge of the lattice at all times. 
It follows from the above that the total intensity $S$ [Eq.~\eqref{eq:total_intenstiy}] is not preserved during the time evolution. 
Nevertheless, the transformed quantity $S_D=\sum_{n}t^n\lvert u_n\rvert^2$
is an integral of motion, such that these amplitudes are bounded as long as the lattice is finite.

Similar stability results are also found for the defocusing case.
The exception here consists of the family of NLSMs arising from the first mode which are always stable as shown by the blue curve in Fig.~\ref{fig:continuation_defocusing_e123}(a).
In fact, these states are known to act as ground states for the chain, since they minimize the energy functional of the system~\cite{RCKG2000}, as is typically
the case in such self-defocusing settings (both continuum and discrete).

We now finish this section with a dynamical observation regarding the choice of perturbation in the results of the linear stability. 
% As we have shown, these vectors are also skinny. 
It is crucial to note that the stability analysis is valid as long as 
we perturb the system with skin-like deviations, $\vec{w} (\tau=0)$ as the ones used in Figs.~\ref{fig:HNdynamics}(a) and (b). 
Without loss of generality, a simple counter example is a perturbation of an otherwise stable NLSM, $\vec{u}$, by a deviation vector $\vec{w} (\tau = 0)=(0,0,\ldots, e^{i\kappa_N})^T$, with $\kappa_n$ being a random phase uniformly drawn in the interval $[-\pi, \pi]$.  
% i.e., using the initial condition $\vec{\psi}(\tau=0) = \vec{u} +\epsilon\vec{w}$.
The time evolution of this perturbed initial condition $\vec{\psi}(\tau=0) = \vec{u} +\epsilon\vec{w}$ is shown in Fig.~\ref{fig:HNdynamics}(c) for the stable NLSM of Fig~\ref{fig:HNdynamics}(a) and Fig.~\ref{fig:eigenvalues_eigenvectors_01a}(a). 
Since this is a stable NLSM, one would expect its dynamics to be straightforwardly
robust, similarly, e.g., to Fig.~\ref{fig:HNdynamics}(a). 
However, for the choice of perturbation $\vec{w}$, 
we observe that the shape of the initial NLSM is greatly modified and after
evolving has moved away from the initial stationary mode. 

To understand the origin of this outcome, for small $\epsilon$, we need to project the perturbation $\vec{w}=(0, 0, \ldots, 1)^T$ into the linear modes of the HN chain, $\vec{w}=\sum_q c_q\vec{u}_q^{(0)}$, following the biorthogonal framework.
It follows that the coefficient $c_q=\vec{v}_q^{T(0)}\vec{w}= t^{-N/2}$ with $q=1, \ldots, N$.
In the example above, we find that $c_q \sim 450$,  such that the perturbation strongly excites all linear modes.
Consequently, the amplitude of $\vec{w}(\tau=0)$ exponentially grows in time, while its center of mass shifts leftward due to the NHSE. 
%When  the amplitude of the perturbation becomes large enough, 
As this perturbation moves to the opposite end of the 
domain where the NLSM is localized,
it interacts with the latter, 
%interacts with the NLSM due to nonlinearity, 
driving the system away from the 
%quasi-periodic 
orbit of the initially stable skin state as seen in Fig.~\ref{fig:HNdynamics}(c) 
as time evolves. Thus, in this nonlinear HN model,
finite perturbations on the ``wrong side'' of the domain
may have detrimental effects through their interaction 
even with stable NLSMs.
%Thus, although the perturbation may
%appear small in the original frame, the above transformatio

\section{\label{sec:concl}Conclusions and Future Challenges}
We have analytically predicted and numerically obtained families of nonlinear skin modes (NLSMs) for the non-reciprocal Hatano-Nelson model in the presence of the Kerr nonlinearity. 
For the case of focusing nonlinearity these modes inherit the property of their linear counterparts and are localized on the preferred side of the lattice. More particularly, in this case, nonlinearity is in synergy with non-reciprocity and it tends to make the modes even more localized than their linear skin counterparts.
Furthermore, we show that the families of NLSMs emerging from the linear limit, terminate at the nonlinear extreme of the anticontinuum limit  where the coupling between sites vanishes.  The analysis of the stability of the modes
near the two limits enables us to characterize their respective stability
characteristics and corresponding eigenvalues. It also allows us to formulate 
a perturbation theory framework near these limits.
For sufficiently strong non-reciprocity we find that the $q$th  family of NLSM will end up in a configuration of $q$ out-of-phase sites in the anticontinuum limit.
These sites may have holes between them if $t$ is sufficiently large, while
they become consecutive for values of $t$ that are below a critical threshold.
Regarding the defocusing case, we have found nonlinear solutions which are still 
skinny but, however, tend to grow in width as the nonlinearity increases. 
The solution in this regime corresponding to $q=1$ is found to always be stable, 
and at high amplitudes it tends to occupy all the (finite) lattice. 
In addition to the existence and stability, we have also explored the
dynamics of the corresponding modes and have shown how the instabilities
(e.g., associated with complex eigenvalue quartets) manifest themselves,
as well as the somewhat unusual (for Hermitian lattices) nature of the impact
of a perturbation at the opposite end to that favored (towards localization)
by the value of the
non-Hermitian parameter $t$.

While we have provided a systematic analysis of the linear and the anti-continuum
modes of the system, various questions still remain to be addressed. 
A detailed analysis of the nonlinear modes from the anticontinuum limit and their 
continuation is worthy of further exploration (cf.~\cite{PKF2005}). 
Dynamically, the modes that emerge from that limit may not only feature
linear (spectral, exponential) instabilities at shorter times, but have
been also shown to manifest nonlinear (power-law) instabilities at 
longer times in the Hermitian case~\cite{avadh}. It would be worthwhile
to explore if such instabilities are present in the Hatano-Nelson model.
Finally, all of our analysis has been performed in the one-dimensional
setting, but such models (in their Hermitian form) are well-known to
have intriguing features, such as (discrete) vortical patterns purely in higher-dimensional
settings~\cite{K2009}, hence the impact of the non-Hermitian nature
on such features would also be another important direction to explore.
Such directions are currently under consideration and will be
presented in future publications.

\begin{acknowledgments}
We express our gratitude to Uwe Thiele for his valuable comments. 
Financial support for V.A.~and B.M.M.~was provided by the {\em Etoiles Montantes en Pays de la Loire} program, within the framework of the NoHENA project.
We acknowledge the support from the US National Science Foundation under 
Grants No.~PHY-2110038 (R.C.G.) and No.~PHY-2110030 (P.G.K.).
\end{acknowledgments}

% \FloatBarrier
\appendix

\section{\label{sec:app:matrices}Dynamical and stability matrices of the Hatano-Nelson model}

The explicit expression of the dynamical matrix of the Hatano-Nelson (HN) lattice model derived from Eq.~\eqref{eq:static_linear_hn_model} with open boundary conditions, i.e. $u_0 = u_{N+1}=0$, yields
\begin{equation}
    H = 
    \begin{pmatrix}
        0 & 1 & 0 & 0 & \ldots & 0 & 0 \\ 
        t & 0 & 1 & 0 & \ldots & 0 & 0 \\
        0 & t & 0 & 1 & \ldots & 0 & 0 \\
        \vdots & \vdots & \vdots & \vdots & \ddots & \vdots & \vdots \\ 
        0 & 0 & 0 & 0 & \ldots & t & 0 \\
    \end{pmatrix},
    \label{eq:Hamiltonian_matrix}
\end{equation}
for unit hopping strength, $C=1$, and parameter of non-reciprocity, $t \in [0, 1)$, between adjacent sites.
% Further, it is also worth providing the expression of the stability matrix.
When nonlinearity is introduced in the HN model, we find the nonlinear solutions $u_n$ at energy $E$ by solving Eq.~\eqref{eq:HNNL_model} and investigate their stability. 
The dynamics of small perturbation, $w_n$ 
%(in the rotating frame $e^{iE\tau}$) 
from a reference nonlinear state, $u_n$ at energy $E$, is generated by its so-called stability matrix,
\begin{widetext}
    \small 
    \begin{equation}
        Z = 
        \left(
        \begin{array}{cccccccccccc}
            E-2g\lvert u_1 \rvert^2 & -C & 0 &  \ldots & 0 & 0 & -g \left(u_1\right)^2 & 0 &  \ldots  & 0 & 0 & 0 \\
            -Ct & E-2g\lvert u_2 \rvert^2 & -C &  \ldots & 0 & 0 & 0 & -g \left(u_2\right)^2  & \ldots  & 0 & 0 & 0 \\
             0 & -Ct &  E-2g\lvert u_3 \rvert^2&  \ldots & 0 & 0 & 0 & 0 &  \ldots  & 0 & 0 & 0 \\
             \vdots & \vdots &\vdots  &\ddots &\vdots &\vdots &\vdots  & \vdots &\ddots  &\vdots &\vdots &\vdots \\
             0 & 0 & 0 &  \ldots & -Ct & E-2g\lvert u_N \rvert^2 & 0 & 0 &  \ldots  & 0 & 0 & -g \left(u_N\right)^2 \\
             g \left(u_1\right)^2 & 0 & 0 &  \ldots & 0 & 0 & -E+2g\lvert u_1 \rvert^2 & C &  \ldots & 0 & 0 & 0 \\
             0 & g \left(u_2\right)^2 & 0 &  \ldots & 0 & 0 & Ct & -E+2g\lvert u_2 \rvert^2 &  \ldots  & 0 & 0 & 0 \\
             \vdots & \vdots &\vdots  &\ddots &\vdots &\vdots &\vdots &\vdots  &\ddots & \vdots &\vdots &\vdots \\
             % 0 & 0 & 0 &  \ldots &g \left(u_{N-1}\right)^2 & 0  & 0 & 0  \ldots & Ct &  -E+2g\lvert u_{N-1} \rvert^2& 0 \\
             0 & 0 & 0 &  \ldots & 0 & g \left(u_N\right)^2 & 0 & 0 &  \ldots & 0 & Ct & -E+2g\lvert u_N \rvert^2 \\
        \end{array}
        \right).
        \label{eq:stabiltiy_matrix}
    \end{equation}
\end{widetext}
The $Z$ is a $2N\times 2N$ matrix, with $N$ denoting the total number of sites of the HN chain.
We rewrite this stability matrix in terms of the product of two operators, $Z=JL$, where
\begin{equation}
    J= 
    \begin{pmatrix}
        0 & I_N \\ 
        -I_N & 0 \\ 
    \end{pmatrix}
\end{equation}
stands for the symplectic matrix, with $I_N$ being the $N\times N$ identity matrix and 
\begin{equation}
    L = J^{-1}Z,
\end{equation}
which depends on the state $u_n$.

\section{\label{sec:app:rpt_linear}Regular perturbation theory: Linear limit}

In this section we perform the RPT in the vicinity of the linear regime ($C=1$) looking at the bifurcation of  the linear modes when the coupling strength $C$ is varied.
We expand the coupling strength as:
\begin{equation}
    C = 1+ \sum_{n} C_n \epsilon^{2n} = 1 + C_1 \epsilon^2 + \mathcal{O}\left(\epsilon^4\right),
\end{equation}
in order to retrieve the linear problem at $\epsilon = 0$.
The choice of this ansatz is based on the fact that by examining the signs of the coefficients $C_n$ gives the direction followed by the bifurcation branch.
% Further, in the first order $C = C_0 + C_1 \epsilon^2$ will gove
In addition, we also expand the solution using
\begin{equation}
    u_n = \sum_k \epsilon^k u_n^{(k-1)} = \epsilon u_n^{(0)} + \epsilon^2 u_n ^{(1)} + \epsilon^3 u_n^{(2)} + \mathcal{O}\left(\epsilon^4\right).
\end{equation}
Thus, as $\epsilon \rightarrow 0$, the amplitude at site  $n$ vanishes.
Substituting these expressions into the time-independent HN equations
\begin{equation}
    E u_n = C\left(u_{n+1} + t u_{n-1} \right) + \lvert u_n \rvert^2 u_n,
\end{equation}
leads to 
\begin{equation}
 \begin{array}{l}
    \epsilon E \left(\vec{u}^{(0)} + \epsilon \vec{u}^{(1)} + \epsilon^2 \vec{u}^{(2)} \right) = \epsilon H \vec{u}^{(0)} 
    \\[2.0ex]
    \quad
    +~\epsilon^2 H\vec{u}^{(1)} + \epsilon^3 \vec{u}^{(2)} + C_1\epsilon^3 \vec{u}^{(0)} 
    \\[2.0ex]
    \quad
    +~\sigma \epsilon^3 \Gamma \left[\vec{u}^{(0)}\right] \vec{u}^{(0)} + \mathcal{O}\left(\epsilon^4 \right),
 \end{array}
\label{eq:perturb_linear_01}
\end{equation}
where $H$ and $\Gamma$ are the dynamical matrix of the HN model and diagonal matrix with non-zero elements $\Gamma_{jj} = \lvert u_j^{(0)}\rvert^2$, respectively. 
Considering $E=E_q$ i.e. the $q$-th eigenvalue of $H$ and collecting terms in $\epsilon$ in Eq.~\eqref{eq:perturb_linear_01}, it follows that for
the first two orders:
\begin{equation}
    \begin{split}
        \epsilon^1 &\colon E_q \vec{u}_q^{(0)} = H\vec{u}_q^{(0)}, \\ 
        \epsilon^2 &\colon E_q \vec{u}_q^{(1)} = H\vec{u}_q^{(1)}. \\
    \end{split}
    \label{eq:perturb_linear_02}
\end{equation}
Eqs.(\ref{eq:perturb_linear_02}) are satisfied if , $\vec{u}_q^{(0)}$ and $\vec{u}_q^{(1)}$ are solutions of the linear HN model.
% , while $\vec{u}_q^{(1)}$ can be seen as their superposition.
Consequently, $\vec{u}_q^{(1)}$ does not contribute to the amplitude correction of the nonlinear solutions.
% of the amplitude and one has to look at higher order terms.
% Further, the solutions of Eqs.~\eqref{eq:perturb_linear_02} are well known, and are the eigenvectors of the linearized lattice.
This correction in amplitude starts to be observed at order $\mathcal{O}(\epsilon^3)$, leading to the non-trivial equations 
\begin{equation}
    E_q \vec{u}_q^{(2)} =  H \vec{u}_q^{(2)} + C_1 H \vec{u}_q^{(0)} + \sigma \Gamma \vec{u}_q^{(0)}.
    \label{eq:perturb_linear_03}
\end{equation}
Multiplying this equation with the left eigenvector $\vec{v}_{q^\prime}$ leads to 
\begin{equation}
    C_1 = -\sigma \frac{\vec{v}^{T(0)} \Gamma \left[\vec{u}_q^{(0)} \right] \vec{u}^{(0)}} {E_q\vec{v}^{T(0)}\vec{u}^{(0)}}    
    \label{eq:perturb_linear_04}
\end{equation}
where we assume the biorthonormalization $\vec{v}_{q^\prime}^{T(0)}\vec{u}_q^{(0)} = \delta_{q,q^\prime}$ for this derivation.
The full expression of $C_1$ is shown in Eq.~\eqref{eq:c1}.
We find that focusing nonlinearity leads to decreasing values of $C$, while the opposite holds for the defocusing case.

\begin{figure}[h!]
    \centering
    \includegraphics[width=0.8\columnwidth]{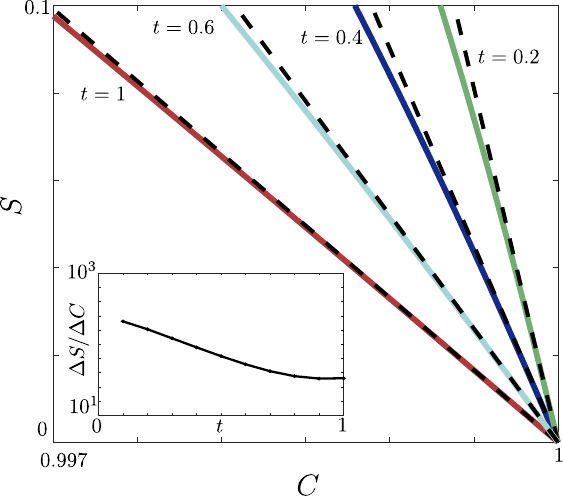}
    \caption{Total intensity against $C$ for families of NLSMs emerging from the first linear mode for four values of $t$ with $t=0.2$, $t=0.4$, $t=0.6$, and $t=1$.
    {Note that in the depicted region (namely, close to the linear limit) all branches 
    correspond to {\em stable} NLSMs.}
    The dashed curves represent the analytical predictions based on the RPT in the neighborhood of the linear skin modes.  
    The inset shows the results of the computation of the rate of growth $\Delta S/\Delta C$ as function of $t$.
    }
    \label{fig:RPT_Linear}
\end{figure}

Figure~\ref{fig:RPT_Linear} shows, in the focusing case $(\sigma=+1$), 
the results of the comparison of the families 
of NLSMs bifurcating from the first linear modes for various $t$ values
obtained using numerical (bold colored curves) and RPT (black curves).
The plot displays the dependence of the total intensity 
as a function of the coupling strengths for these families and shows  a good agreement between the two results for small amplitude waves.
As expected, the deviation (slowly) increases as we depart from the unit coupling limit,
given the strong dependence of $S$ on $C$.
Furthermore, we also find that the amplitude of these weakly nonlinear states grows faster for strongly non-reciprocal chains compared to the ones with $t\rightarrow 1$; see the inset of Fig.~\ref{fig:RPT_Linear}.

%%%%%%%%%%%%%%%%%%%%%%%%%%%%%%%%%%%%%%%%%%%%%%%%%%%%%%%%%%%%%%%%%%%
\begin{figure}[h!]
    \centering
    \includegraphics[width=0.9\columnwidth]{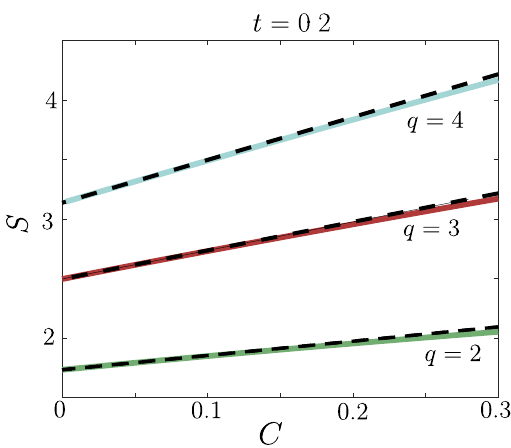}
    \caption{
    Same as in Fig.~\ref{fig:RPT_Linear}, but for families emerging, near
    the AC (rather than the linear) limit, from the second ($q=2$), 
    third ($q=3$) and fourth ($q=4$) linear skin modes with $t=0.2$.
       {Note that in the depicted region (namely, close to the anticontinuum limit) all branches 
    correspond to {\em stable} NLSMs.}
    The dashed lines represent the analytical predictions based on the RPT in the vicinity of the AC limit.  
    % The inset is the results of the computation of the rate of growth $\Delta S/\Delta C$ as function of $t$.
    }
    \label{fig:RPT_AClimit}
\end{figure}
%%%%%%%%%%%%%%%%%%%%%%%%%%%%%%%%%%%%%%%%%%%%%%%%%%%%%%%%%%%%%%%%%%%

%%%%%%%%%%%%%%%%%%%%%%%%%%%%%%%%%%%%%%%%%%%%%%%%%%%%%%%%%%%%%%%%%%%
\section{\label{sec:app:RPTAClimit}Regular perturbation theory: Anticontinuum limit}
%%%%%%%%%%%%%%%%%%%%%%%%%%%%%%%%%%%%%%%%%%%%%%%%%%%%%%%%%%%%%%%%%%%

We now apply the RPT to show the existence and stability of high amplitude NLSMs when the coupling strength is varied from the AC ($C=0$) limit~\cite{M2018,PKF2005,K2009}. 
Without loss of generality, we focus here on the so-called lower order solutions, which arise from initially compact (localized) states of arbitrary size (width) with $M>2$ at the AC limit.
Such NLSMs belong to the families emerging from the linear modes with $t<t_c$, in line with what was discussed in the main text.
Assuming the ansatz,
\begin{equation}
    \begin{split}
        C &= \beta + \mathcal{O}(\beta^2), \\
        u_n &=\tilde{u}_n^{(0)} + \beta \tilde{u}_n^{(1)} + \mathcal{O}(\beta^2),
    \end{split}
    \label{eq:ansatz_aclimit}
\end{equation}
where $\beta \ll 1$, and substituting it into the time-independent nonlinear HN equations
\begin{equation}
    E u_n = C\left(u_{n+1} + t u_{n-1} \right) + \lvert u_n \rvert^2 u_n,
\end{equation}
we obtain, 
\begin{equation}
    \nonumber
    \begin{array}{l}
    E \left(\tilde{u}_n^{(0)} + \beta \tilde{u}_n^{(1)} \right) = \beta \left( \tilde{u}_{n+1}^{(0)} + t \tilde{u}_{n-1}^{(0)} + \beta \tilde{u}_{n+1}^{(1)} + t\beta \tilde{u}_n^{(1)}\right) 
    \\[2.0ex]
    \quad
    +~\left[\lvert \tilde{u}_n^{(0)}\rvert^2 + 2\beta \tilde{u}_n^{(0)} \tilde{u}_n^{(1)} \right] \left[\tilde{u}_n^{(0)} + \beta \tilde{u}_n^{(1)} \right]    + \mathcal{O}\left(\beta^2\right).
    \end{array}
\end{equation}
Neglecting the $\mathcal{O}\left(\beta^2\right)$ and collecting the terms in series of $\beta$ leads to 
\begin{equation}
    E\tilde{u}_n^{(0)} = \lvert \tilde{u}_n^{(0)} \rvert^2 \tilde{u}_n^{(0)},
    \label{eq:perturb_aclimit_01}
\end{equation}
at order $\mathcal{O}(\beta^0)$.
These constitute the discrete equations of the chain at the AC limit 
for which oscillators are independent. 
As such, the stationary solutions are well known
\begin{equation}
    \tilde{u}_n^{(0)} = \sqrt{E}e^{i\theta_n},
\end{equation}
for the excited (finitely many) 
sites, while at this leading order the remaining nodes have $\tilde{u}_n^{(0)}=0$.
Furthermore, since we focus on real solutions,  we impose the phase $\theta_n = 0,~\pi$, which automatically satisfies the corresponding (mass) flux conditions~\cite{PKF2005}.
%The latter being of outmost importance for the rest of the derivation.

For the order $\mathcal{O}(\beta^1)$, we have 
\begin{equation}
    E \tilde{u}_n^{(1)} = \left(\tilde{u}_{n+1}^{(0)} + t \tilde{u}_{n-1}^{(0)} \right) + 3\lvert \tilde{u}_n^{(0)}\rvert^2 \tilde{u}_n^{(1)}.
\end{equation}
This expression further reduces to 
\begin{equation}
    -2E \tilde{u}_n^{(1)} = \left(\tilde{u}_{n+1}^{(0)} + t \tilde{u}_{n-1}^{(0)} \right),
    \label{eq:pertub_02}
\end{equation}
when considering Eq.~\eqref{eq:perturb_aclimit_01}.
Since we are interested in NLSMs, we can assume that the initial guess at the AC limit is located at the left edge of the chain.
Consequently, we write the explicit expression of the first order correction in amplitude:  
\begin{equation}
    \begin{split}
    \tilde{u}_1^{(1)} &= -\frac{e^{-i\theta_1}}{2\sqrt{E}}\cos \left(\theta_{2} - \theta_1 \right), \\ 
        \tilde{u}_n^{(1)} &= -\frac{e^{-i\theta_n}}{2\sqrt{E}} \left[\cos \left(\theta_{n+1} - \theta_n \right) + t\cos \left(\theta_{n-1} - \theta_n \right) \right], \\ 
        \tilde{u}_M^{(1)} &= -\frac{e^{-i\theta_M}}{2\sqrt{E}}  t\cos \left(\theta_{M-1} - \theta_M \right), 
    \end{split}
    \nonumber
    %\label{eq:un_RPT}
\end{equation}
with $n=2, \ldots, M-1$.
% Here we considered that $e^{i\Delta \theta} = \cos \Delta \theta$, if $\Delta\theta = 0,~\pi$ and that we are starting from an $L$-site compact stationary solution from the AC limit.
Clearly, the amplitude of the first order correction is $\tilde{u}_n^{(1)}$ and it 
is inversely proportional to $\sqrt{E}$.
On the other hand, the presence of non-reciprocity through $t$, induces a symmetry breaking at finite coupling.

Having obtained the different high amplitude NLSMs near the AC limit,
we further proceed to characterize their stability.
This is done by calculating the largest eigenvalues $\lambda$ of the stability 
matrix, $Z$ [Eq.~\eqref{eq:stabiltiy_matrix}] using as reference states 
the nonlinear states of the RPT above. This yields
\begin{equation}
    \lambda^2 = 2E\beta \gamma^{(1)}, 
    \label{eq:eigenvalues_AClimit}
\end{equation}
with $\gamma^{(1)}$ being the first order correction of the largest eigenvalues of $M$ from the AC limit which depends on the number of excited oscillators, $M$.
Here, we report the results for the two- ($M=2$) and three- ($M=3$) site NLSMs.
It follows that for $M=2$,
\begin{equation}
   \gamma^{(1)} = (1 + t)c_1,
   \label{eq:eigenvalues_gamma_M2}
\end{equation}
while for the case  with $M=3$,
\begin{eqnarray}
\gamma_{\pm}^{(1)} &=& \frac{1}{2} \bigg\{ (1+t)(c_1+ c_2)
\\[1.0ex] 
\nonumber
&&
\left.
\pm \sqrt{(1+t)^2 \left[c_1^2 + c_2^2 \right] - 2 (1+t^2) 
c_1 c_2 } \right\}
       \label{eq:eigenvalues_gamma_M3}
\end{eqnarray}
where $c_n\equiv  \cos \Delta \theta_n$ and
$\Delta \theta_n = \theta_{n}- \theta_{n-1}$.
For the families arising from the linear modes we have $\Delta \theta_n = \pm \pi$ which, in turn, can be seen to lead
these high amplitude NLSMs to be generically stable near
the AC-limit, similarly to their Hermitian 
counterparts~\cite{K2009}.

Figure~\ref{fig:RPT_AClimit} shows the dependence of the total intensity on the coupling strength for the families emerging from the second, third and forth linear skin modes at $t=0.2$.
In the neighborhood of the AC limit, it is clear that there is good agreement between the RPT (black curves) and the numerical continuations (colored curves).
Furthermore, the stability of these high intensity NLSMs calculated from the numerical simulations clearly demonstrates that these states are  linearly stable (see colored curves in Fig.~\ref{fig:RPT_AClimit}).

% The \nocite command causes all entries in a bibliography to be printed out
% whether or not they are actually referenced in the text. This is appropriate
% for the sample file to show the different styles of references, but authors
% most likely will not want to use it.
%\nocite{*}
% \bibliographystyle{apsrev4}
\let\itshape\upshape
\normalem

\bibliography{references}% Produces the bibliography via BibTeX.

\begin{thebibliography}{56}
\bibitem [{BIF()}]{BIFURCT1}%
  \BibitemOpen
  \href@noop {} {}\bibinfo {note} {For the DNLS model (HN system with $t=1$),
  we mention the presence of a Pitchfork bifurcation point located at
  $C=0.9036454$. At this juncture, the branch stemming from the normal mode
  splits into three branches. Two of these branches converge to the one-site
  solutions at positions $n=N/2$ and $n=N/2+1$ along the chain in the AC limit,
  with $N$ being the size of the chain. The remaining branch reaches the
  two-site solution at $C=0$. In case of the Hatano-Nelson lattice ($t\neq 1$),
  this Pitchfork bifurcation point undergoes a transformation into a limit
  point which effectively segregate these families of solutions.}\BibitemShut
  {Stop}%
\end{thebibliography}

\end{document}